\documentclass[aps,reprint, prx, superscriptaddress, showpacs,nobibnotes]{revtex4-1}
\usepackage{graphicx}
\usepackage{bm,amsmath}
\usepackage{systeme}
\usepackage{hyperref}
\usepackage{mathrsfs}
\usepackage{dcolumn}
\usepackage{natbib}
\usepackage[dvipsnames]{xcolor}
\usepackage{ulem}

\usepackage[version=4]{mhchem}
\usepackage[title]{appendix}
\begin{document}
\title{Disorder-induced local strain distribution in Y-substituted \ce{TmVO4}}

\author{Yuntian Li}
\affiliation{Department of Applied Physics, Stanford University, Stanford, California 94305, USA}
\affiliation{Geballe Laboratory for Advanced Materials, Stanford University, Stanford, California 94305, USA}
\author{Mark P. Zic}
\affiliation{Geballe Laboratory for Advanced Materials, Stanford University, Stanford, California 94305, USA}
\affiliation{Department of Physics, Stanford University, Stanford, California 94305, USA}
\author{Linda Ye}
\affiliation{Department of Applied Physics, Stanford University, Stanford, California 94305, USA}
\affiliation{Geballe Laboratory for Advanced Materials, Stanford University, Stanford, California 94305, USA}
\author{W. Joe Meese}
\affiliation{School of Physics and Astronomy, University of Minnesota, Minneapolis, Minnesota 55455, USA}
\author{Pierre Massat}
\affiliation{Geballe Laboratory for Advanced Materials, Stanford University, Stanford, California 94305, USA}
\affiliation{Department of Physics, Stanford University, Stanford, California 94305, USA}
\author{Yanbing Zhu}
\affiliation{Department of Applied Physics, Stanford University, Stanford, California 94305, USA}
\author{Rafael M. Fernandes}
\affiliation{School of Physics and Astronomy, University of Minnesota, Minneapolis, Minnesota 55455, USA}

\author{Ian R. Fisher}
\affiliation{Department of Applied Physics, Stanford University, Stanford, California 94305, USA}
\affiliation{Geballe Laboratory for Advanced Materials, Stanford University, Stanford, California 94305, USA}

\date{\today}

\newcommand{\JoeColor}{BurntOrange}
\newcommand\Joesout{\bgroup\markoverwith{\textcolor{\JoeColor}{\rule[0.5ex]{4pt}{2pt}}}\ULon}
\newcommand{\JoeInsert}[1]{{\color{\JoeColor} #1}}
\newcommand{\JoeComment}[1]{\JoeInsert{(Joe Comment): #1}}
\newcommand{\JoeReplacement}[2]{\JoeInsert{#1} \Joesout{#2}}

\newcommand{\YuntianColor}{Magenta}
\newcommand{\YuntianInsert}[1]{{\color{\YuntianColor} #1}}
\newcommand{\YuntianComment}[1]{\YuntianInsert{(Joe Comment): #1}}
\newcommand{\YuntianReplacement}[2]{\YuntianInsert{#1} \Yuntiansout{#2}}

\begin{abstract}
We report an investigation of the effect of substitution of Y for Tm in \ce{Tm_{1-x}Y_xVO4} via low-temperature heat capacity measurements, with the yttrium content $x$ varying from $0$ to $0.997$. Because the Tm ions support a local quadrupolar (nematic) moment, they act as reporters of the local strain state in the material, with the splitting of the ion's non-Kramers crystal field groundstate proportional to the quadrature sum of the in-plane tetragonal symmetry-breaking transverse and longitudinal strains experienced by each ion individually. Analysis of the heat capacity therefore provides detailed insights into the distribution of local strains that arise as a consequence of the chemical substitution. These local strains suppress long-range quadrupole order for $x>0.22$, and result in a broad Schottky-like feature for higher concentrations. Heat capacity data are compared to expectations for a distribution of uncorrelated (random) strains. For dilute Tm concentrations, the heat capacity cannot be accounted for by randomly distributed strains, demonstrating the presence of significant strain correlations between sites. For intermediate Tm concentrations, these correlations must still exist, but the data cannot be distinguished from that which would be obtained from a 2D Gaussian distribution. The cross-over between these limits is discussed in terms of the interplay of key length scales in the substituted material. The central result of this work, that local strains arising from chemical substitution are not uncorrelated, has implications for the range of validity of theoretical models based on random effective fields that are used to describe such chemically substituted materials, particularly when electronic nematic correlations are present.      


\end{abstract}

\maketitle

\section{\NoCaseChange{Introduction}}
Phase diagrams of a wide variety of strongly correlated electron systems can be traversed via chemical substitution. The tuning effect of different dopants in different materials can be attributed to a variety of physical effects. However, a pervasive and inescapable effect of such substitution must always be the associated inhomogeneous internal strains which are induced by incorporating inequivalent atoms to a crystal lattice. These local strains can have a profound effect on various types of emergent electronic order and phase transitions. Here we focus specifically on the case of electronic nematic order, whose bilinear coupling with strains of the same symmetry leads to a particularly strong interplay. The bilinear coupling means that strain plays the role of an effective conjugate field for the nematic order parameter, and hence the inhomogeneous strains which arise from chemical substitution have been conventionally understood as a random field \cite{TVojta_Disorder_Review} both experimentally \cite{Canfield_Annealing_2011, Curro_NMR_2016,Birgeneau_Annealing_2016} and theoretically \cite{Carlson_Hysteresis_2006,Carlson_Decoding_2014,Nie_Kivelson_2014,TVojta_Stripes_2022,Meese2022, Schmalian_qEA_2021}. Despite this appealing perspective, strains in solids relax over long distances, implying a significant correlation between adjacent sites \cite{landauTheoryElasticity1970, muraMicromechanics}. Similarly, the elastic response of solids (described by the elastic stiffness tensor) means that local strains of different symmetry even on a specific site are not necessarily uncorrelated. These simple perspectives imply that the degree to which a description of the effects of inhomogeneous strains arising from chemical substitution as a truly random effect should be examined in some detail. Indeed, a characterization of ``randomness" associated with the long-range effects of local strains in a real material system, and the applicability of the random field assumption, remain respectively an open challenge and an open question.


\begin{figure}[ht!]
	\includegraphics[width = \columnwidth]{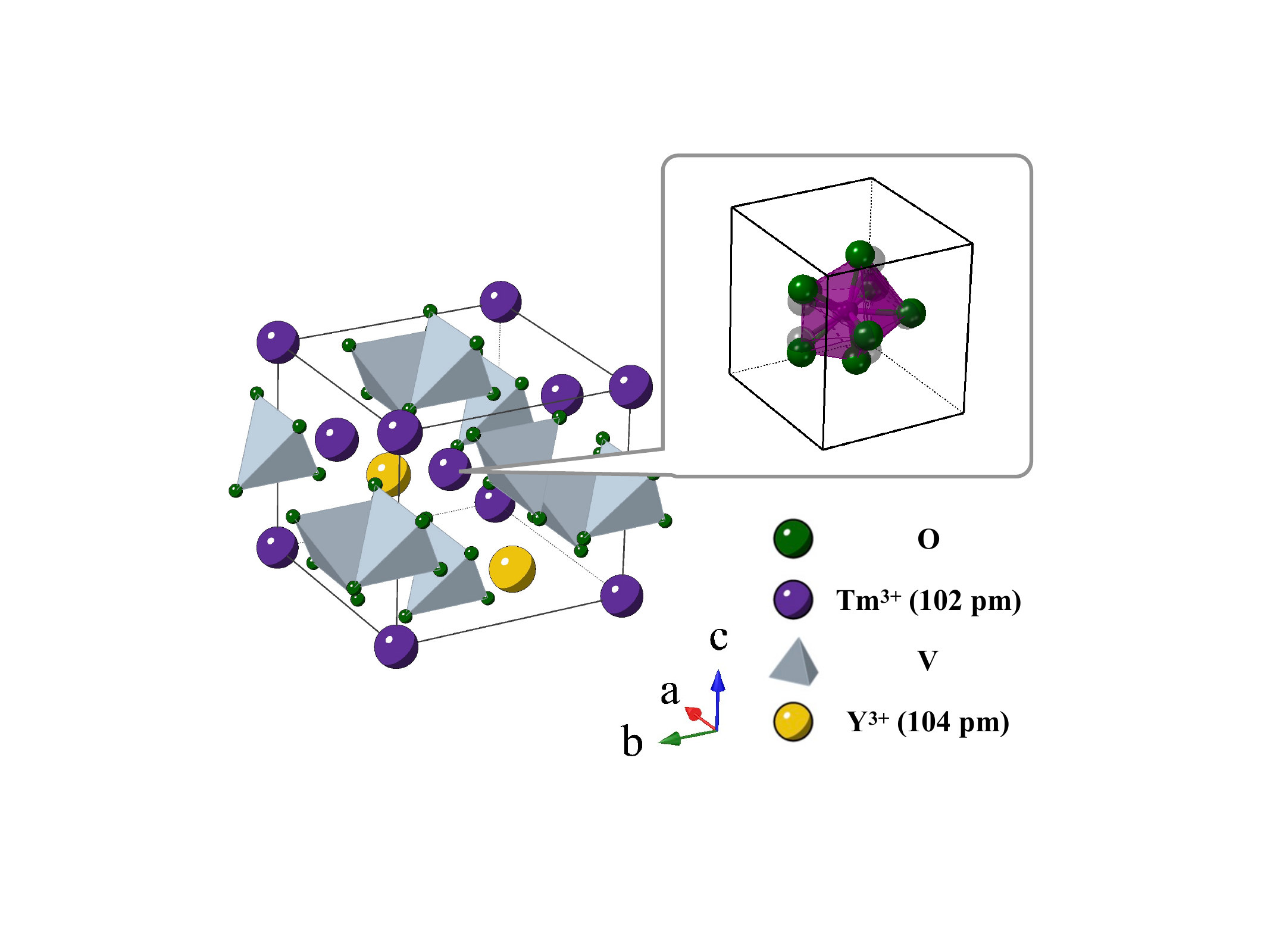}
	\caption{\label{fig-1} Crystal structure of the hypothetical material \ce{Y1Tm3V4O16} obtained via density functional theory (DFT) calculations, with \ce{Tm} (\ce{Y}) atoms represented by purple (yellow) spheres. \ce{VO4} units are shown as gray coordination polyhedra, with \ce{O} atoms as green spheres. Inset shows a magnified view of a distorted \ce{TmO8} polyhedron overlaid with that of pristine \ce{TmVO4} (\ce{O} atoms in the pristine structure are shown as grey spheres) to illustrate the distortions induced by the Y substitution The atomic displacements of the \ce{O} sites in the inset are magnified 40 times for clarity. Ionic radii of the \ce{Tm^{3+}} and \ce{Y^{3+}} ions are indicated. }
\end{figure}

In this work, we attempt to experimentally address the above issues in a model system exhibiting ferroquadrupolar order, which may be viewed as a simple local-moment realization of Ising nematic order \cite{maharaj2017transverse,tmvo42022}). The subject of this study, \ce{TmVO4}, is an insulator with a tetragonal zircon-type \ce{RVO4} crystal structure. The crystal electric field (CEF) ground state of the \ce{Tm^{3+}} ions is a non-Kramers doublet, with the first excited CEF state well above the doublet in energy ($77$ K, $54 \mathrm{cm}^{-1}$) \cite{Morin1990}, rendering \ce{TmVO4} a good approximation to an ideal pseudospin-$\frac{1}{2}$ system at low-temperature \cite{wortman1974analysis}. 

\ce{TmVO4} is known to host considerable quadrupole-lattice interactions, resulting in complete softening of the $c_{66}$ elastic modulus and a continuous cooperative Jahn-Teller phase transition at $T_Q=2.2$ K \cite{MELCHER1976,tmvo42022}. The Tm ions in the low-temperature phase develop a spontaneous local quadrupole moment (splitting the non-Kramers Jahn-Teller-active CEF groundstate) while the lattice develops a uniform strain $\epsilon_{xy}$ of the same symmetry. We introduce disorder to \ce{TmVO4} by substituting Tm with Y.  Aside from the overall dilution effect of the magnetic species and quadrupole moments, the substitution of \ce{Y} introduces local strains that can couple to the ferroquadrupolar (Ising-nematic) order parameter. To illustrate the anticipated effect of substitution, Fig.\ref{fig-1} 
shows a schematic of the local distorted crystal field environments of the \ce{Tm^{3+}} ion obtained from Density Functional Theory (DFT) calculations of a hypothetical \ce{Y1Tm3V4O16} ordered super-cell (investigated for computational simplicity). Inspection of the figure reveals how \ce{Y} sites distort nearby \ce{TmO8} clusters, resulting in a change of the local crystal field environment for the Tm$^{3+}$ ions relative to the case of \ce{TmVO4}. While substitution in \ce{Tm_{1-x}Y_x VO4} does not result in an ordered superlattice, the effect of chemical substitution in terms of inducing local strains on nearby Tm ions will be comparable. It is the effect of these substitution-induced strains that we investigate in the present study. 

Noting that the non-Kramers doublet can be split by both $\epsilon_{x^2-y^2}$ and $\epsilon_{xy}$ strains (which transform as the $B_{1g}$ and $B_{2g}$ irreducible representations of the tetragonal point group $D_{4h}$, respectively) \cite{maharaj2017transverse}, these two independent symmetry channels naturally span a two-dimensional space. According to the central limit theorem, uncorrelated random strains in the pseudo-spin space follow a Gaussian distribution, with their means centered at zero (Fig. \ref{fig-2}(a)). The total splitting of the non-Kramers doublet $\Delta$ is determined by the quadrature sum of the strain field in the two-symmetry channels $\sqrt{(\eta_1\epsilon_{x^2-y^2})^2+(\eta_2\epsilon_{xy})^2}$, where $\eta_1$ and $\eta_2$ are nemato-elastic coupling strengths (Fig. \ref{fig-2}(b)). As a result, the probability distribution of $\Delta$ can be obtained from joint probability density functions (PDF) in both dimensions (Fig. \ref{fig-2}(c)). We will return to the exact functional form of the distribution in the case of uncorrelated local strains (2D-Gaussian) below. 

Experimentally, $\Delta$ is accessible through thermodynamic measurements. In the following, we map out the entropy landscape of $4f$ quadrupoles in \ce{Tm_{1-x}Y_{x}VO4} ($x=0\sim1$) with heat capacity experiments, from which we extract the evolution of local strain distributions with Tm concentration, to shed light on the correlation between local strains.

A well-known minimal theoretical model that captures the effect of quenched random fields on phase transitions is the random field Ising model (RFIM) \cite{Imry_Ma,Aharony1976,Grinstein1976}, in which even weak random fields will suppress second-order phase transitions to a lower temperature for systems with either short-ranged or long-ranged interactions \cite{TVojta_Disorder_Review,Natterman_Review_1998,Binder_RFIM_1983,Gehring_MFRFIM_1976}. Indeed, in three-dimensional systems, there exists a threshold random field strength at which the phase transition will be fully suppressed, whereas in two dimensions long-range order is completely eliminated even at zero temperature \cite{Zachar2003}. Due to its conceptual simplicity, the RFIM continues to be an important model in understanding the phase behavior of electronic nematics of real materials in the presence of structural disorder \cite{Carlson_Hysteresis_2006,Carlson_Decoding_2014,Nie_Kivelson_2014,TVojta_Stripes_2022,Meese2022, Schmalian_qEA_2021}. .

An important definition one must include when applying the RFIM, however, is ``randomness". In other words, the characterization of a distribution of random fields that are coupled to the order parameter. This is somewhat arbitrary since the only distribution moment that can be clearly seen by experimental probes is the first, i.e. the mean value of the random field. Higher-order moments, encoding the spatial correlation in the random fields, can be postulated \textit{ab initio} \cite{TVojta_Disorder_Review,Natterman_Review_1998}. A frequent choice, however, is to ignore spatial correlation entirely, and assume furthermore that the random fields are all identically drawn from a Gaussian distribution with zero mean and some standard deviation which characterizes the ``disorder strength". This choice is common for theoretical treatments of electronic nematic systems as well, but due to the long-range nature of elasticity, it is unclear whether one can treat the random fields as uncorrelated without losing essential physics in actual materials. Our study explores evidence for the presence of these correlations as a first step towards assessing their significance.

\begin{figure}[tp]
	\includegraphics[width = \columnwidth]{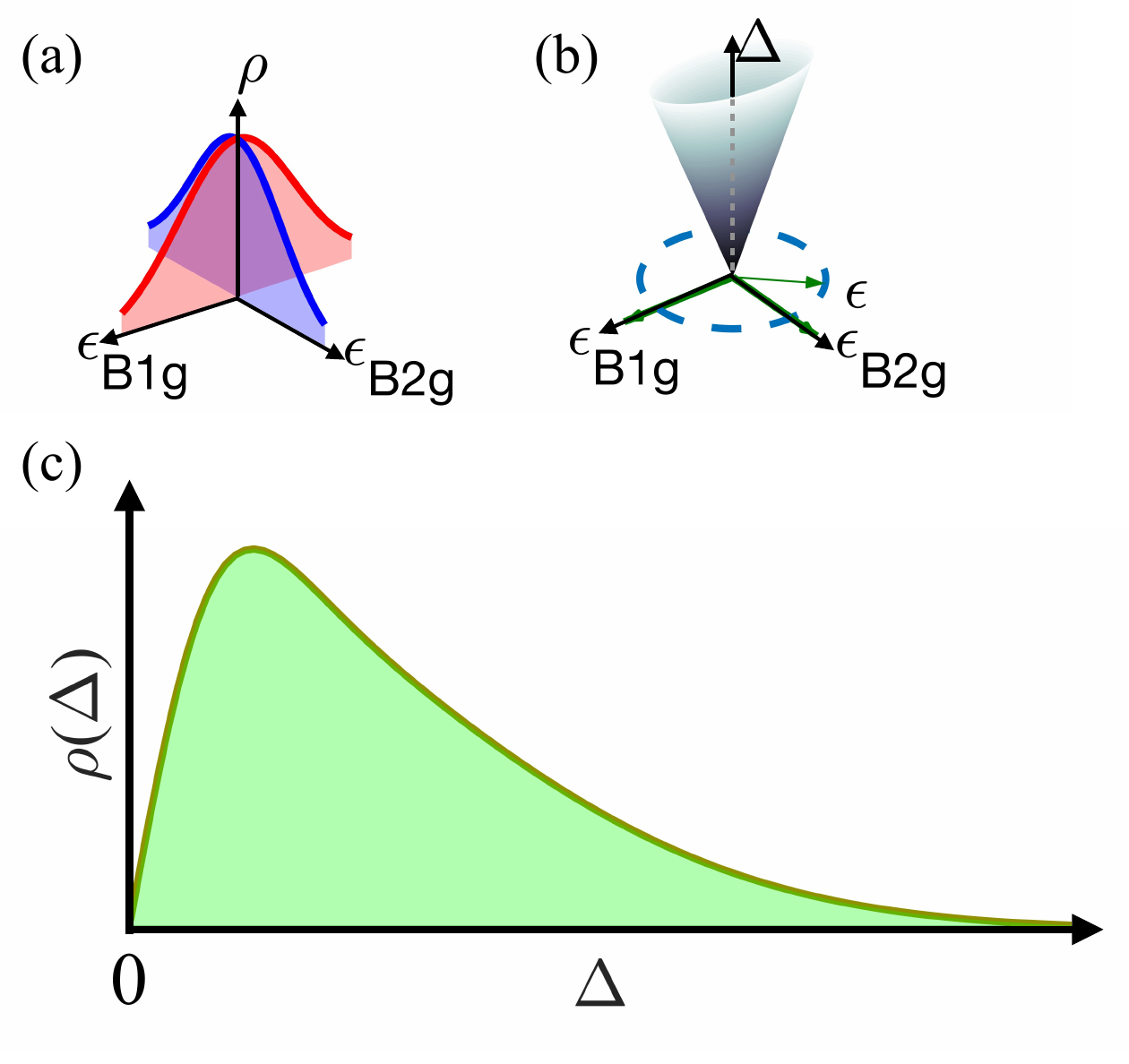}  
	\caption{(a) Schematic of the two-dimensional strain distribution $\rho$ as a function of strain with $B_{1g}$ ($\epsilon_{B_{1g}}=\epsilon_{x^2-y^2}=\frac{1}{2}(\epsilon_{xx}-\epsilon_{yy})$) and $B_{2g}$ ($\epsilon_{B_{2g}}=\epsilon_{xy}$) symmetry, expected from the central limit theorem for uncorrelated strains wherein each of them has a functional form of a Gaussian distribution centered at zero. (b) The magnitude of the local strain-induced energy gap, $\Delta$, measured in arbitrary units, as the quadrature sum of the induced gap in both $\epsilon_{B_{1g}}$ and $\epsilon_{B_{2g}}$ channels. The color scale follows Eq. \ref{eq:2D_Gaussian_eqn}. (c) The resulting probability distribution $\rho(\Delta)$ from the double-Gaussian distribution (see section IV). \label{fig-2}}
\end{figure}

\section{\NoCaseChange{Experimental Methods}}

Single crystals of \ce{Tm_{1-x}Y_{x}VO4} used in the heat capacity measurements were grown via slow cooling in a flux of \ce{Pb2V2O7} using a mixture of \ce{Tm2O3} and \ce{Y2O3} precursors. More details related to material synthesis can be found in Refs. \cite{feigelson1968flux,smith1974flux,oka2006crystal}. The \ce{Y} content $x$ of each sample is further determined by two different methods: for $x\leq0.315$, the composition is determined by microprobe analysis carried out at the Stanford Mineral and Microchemical Analysis Facility; for $x>0.315$, the \ce{Tm} composition is determined by the magnetic moment extracted from a.c. magnetic susceptibility measurements in a Quantum Design Magnetic Property Measurement System (MPMS) in a zero DC magnetic field. More details of the determination of the doping with the MPMS measurement can be found in Appendix \ref{app:chem_composition}.

The high-temperature heat capacity of \ce{Tm_{1-x}Y_{x}VO4} (4 K-35 K) was measured in a Quantum Design Physical Property Measurement System (PPMS) under high vacuum conditions with the thermal relaxation technique. The low-temperature heat capacity measurement was carried out with the same method in a dilution refrigerator insert from 4 K down to the base temperature. All data were taken at a zero external magnetic field. The data analysis is performed with the nonlinear least-squares fitting method.

The super-cell in Fig.1 was created using VESTA from the crystal structure of \ce{TmVO4} obtained from the Crystallographic Information File (CIF) in the Cambridge Crystallographic Data Center (CCDC) (accession code 2117139)\cite{tmvo42022}. DFT calculations for the ionic relaxation were performed using the generalized gradient approximation (GGA) exchange-correlation functional of Perdew, Burke, and Ernzerhof (PBE) \cite{Perdew1996} as implemented in the Vienna Ab initio Simulation Package (VASP) \cite{Kresse1996}. The projected augmented wave (PAW) method \cite{Bloch1994,Kresse1999} was used for ion–electron interaction, and the conjugate-gradient algorithm (IBRION = $2$) was used to relax the ions into their instantaneous ground state.

\section{\NoCaseChange{Heat Capacity of \ce{Tm_{1-x}Y_{x}VO4}}}

In Fig. \ref{fig-3} we show the heat capacity of \ce{Tm_{1-x}Y_{x}VO4} from $x=0$ to $x=0.997$. Panel (a) shows the measured total heat capacity, $C_p$, normalized by the molar gas constant $R$, while panels (b) and (c) show the derived $4f$ contribution to the heat capacity, $C_p^{4f}$,  (the isolation of $C_p^{4f}$ from the phonon contributions to $C_p$ is described in detail in Appendix \ref{app:specific_heat_determination}). The ferroquadrupolar transition temperature $T_Q$, defined here by taking the average of the maxima and minima of the second derivative of $C_p$, is highlighted with a red triangle for each composition in Fig. \ref{fig-3}(a). The phase transition is fully suppressed near $x=0.22$. We note that $T_Q$ is suppressed considerably more rapidly than what would be expected based solely on the dilution effects of the magnetic ions. This result, even before an entropy analysis, already suggests that local strains induced by Y-substitution play a key role in the suppression of $T_Q$.


Inspection of Fig. \ref{fig-3}(a) reveals the presence of a broad shoulder-like feature in the heat capacity for temperatures above $T_Q$ for compositions above $x=0.15$ (cyan curve in Fig. \ref{fig-3}(a)). This shoulder becomes progressively more prominent with increasing $x$ as the phase transition is fully suppressed (Fig. \ref{fig-3}(b)). This pervasive feature can be attributed to the Schottky anomaly associated with the strain-induced splitting of the ground state crystal field doublet of \ce{Tm^{3+}}. However, the anomaly is never described by a single gap -- rather, there is a distribution of gaps. To illustrate this point, in Fig. \ref{fig-3}(b), we plot $C_p^{4f}$ normalized with its maximum $C_p^{4f,max}$  against $T$ normalized by $T_{0}$ ($T_{0}$ refers to the temperature at which $C_p^{4f}$ is maximized). For comparison, we include in this figure a trace (black dashed line) showing the Schottky anomaly function that would arise from a single gap $\Delta$. The strong deviation of the observed $C_p^{4f}/C_p^{4f,max}$ clearly indicates that there exists a finite distribution of $\Delta$, which originates from the anticipated inhomogeneous distribution of substitution-induced local strains. Furthermore, it is clear from inspection of Fig. \ref{fig-3}(b) that the distribution of strain-induced gaps must change as a function of composition (since the overall shape of the normalized curves changes as a function of composition). Analysis of this variation informs the key insights of the present work.  



The $T$-evolution of $C_p^{4f}/T$ is shown in Fig. \ref{fig-3}(c) for all compositions, providing the raw data from which the integrated entropy can be extracted. In Fig. \ref{fig-4}(a), we extract the total entropy associated with the $4f$ degree of freedom via numerical integration $S^{4f}=\sum (C_p/T) \Delta T$. The $\Delta S^{4f}$ plotted here represents the release of entropy with respect to that at high temperature: $\Delta S^{4f}(T)=S^{4f}(T)-S^{4f}\mid_{T=T_{max}}$. We then plot the release of the entropy with respect to the saturated value at high $T$, from the lowest accessible $T$. For most compositions (all but the very dilute Tm concentrations), a large fraction of the total entropy $k_B\ln(2)$ per Tm is recovered upon cooling from 4 K down to below 100 mK, indicating that nearly all  \ce{Tm^{3+}} experience a significant local strain-induced splitting. The entropy landscape in the $x-T$ plane shown in Fig. \ref{fig-4}(b) also highlights that beyond $x=0.22$ where $T_Q$ is fully suppressed, the temperature scale associated with the entropy release first increases and becomes comparable with $T_Q$ of the pristine \ce{TmVO4} near $x=0.5$, before progressively decreasing towards 0 in the dilute Tm limit. Such energy scales are consistent with the large magnetoelastic coupling strength experienced by the non-Kramers doublet ground state of \ce{Tm^{3+}}, as well as the rapid suppression of $T_Q$ with $x$.

\begin{figure}[t]
	\includegraphics[width = \columnwidth]{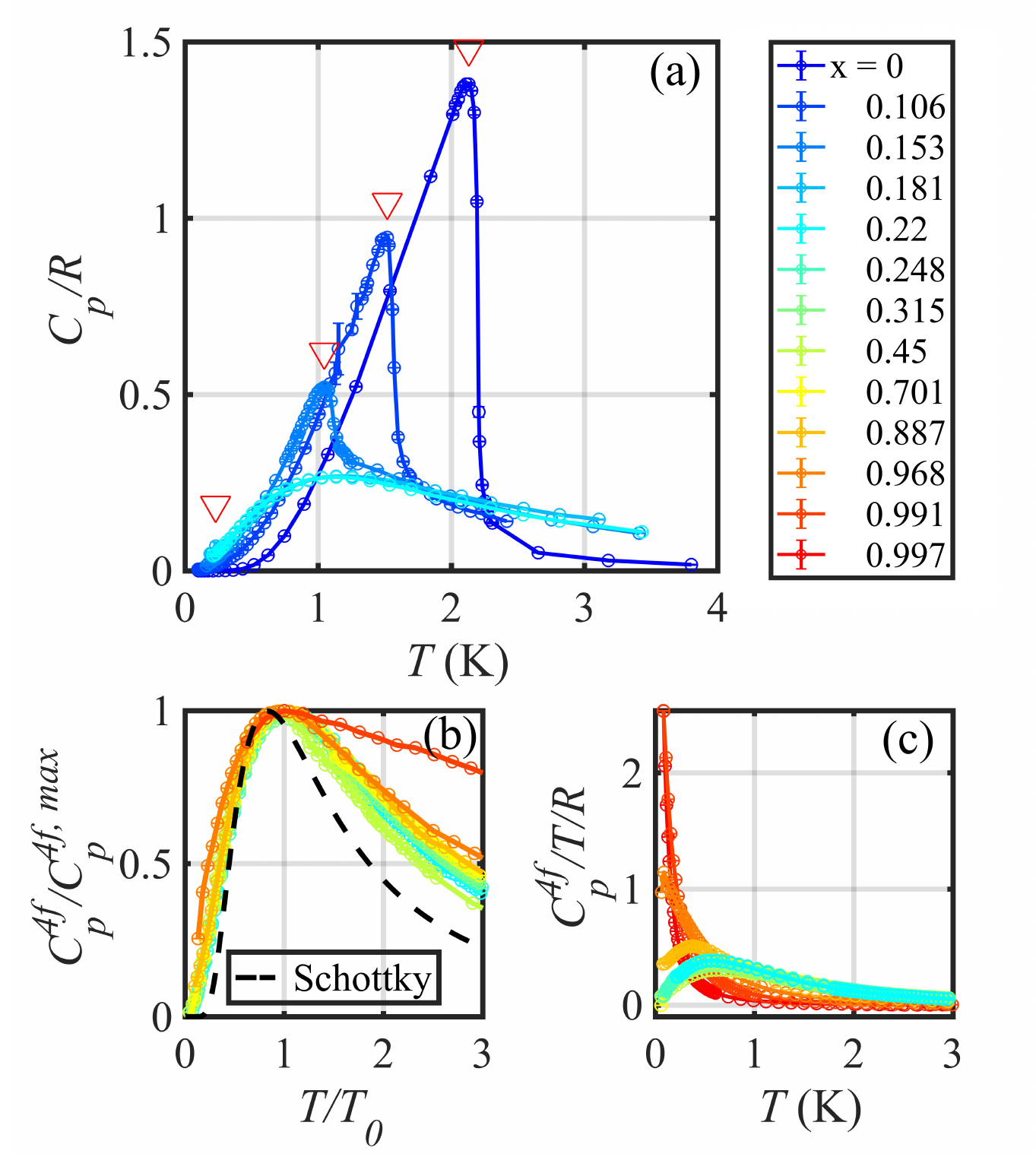}
	\caption{(a) Temperature and chemical composition dependence of molar heat capacity $C_p$ (normalized by $R$) of \ce{Tm_{1-x}Y_{x}VO4} for $x\leq0.22$. The peak in $C_p$ that corresponds to the ferroquadrupolar phase transition is labeled by red arrows.  (b) $C_P^{4f}$ normalized by the maximum $C_p^{max}$ with respect to $T$ normalized by $T_0$ (see text) for $x>0.22$. (c) $T$-evolution of $C_p^{4f}/T$ for $x>0.22$.} \label{fig-3} 
\end{figure}

\begin{figure}[t]
	\includegraphics[width = \columnwidth]{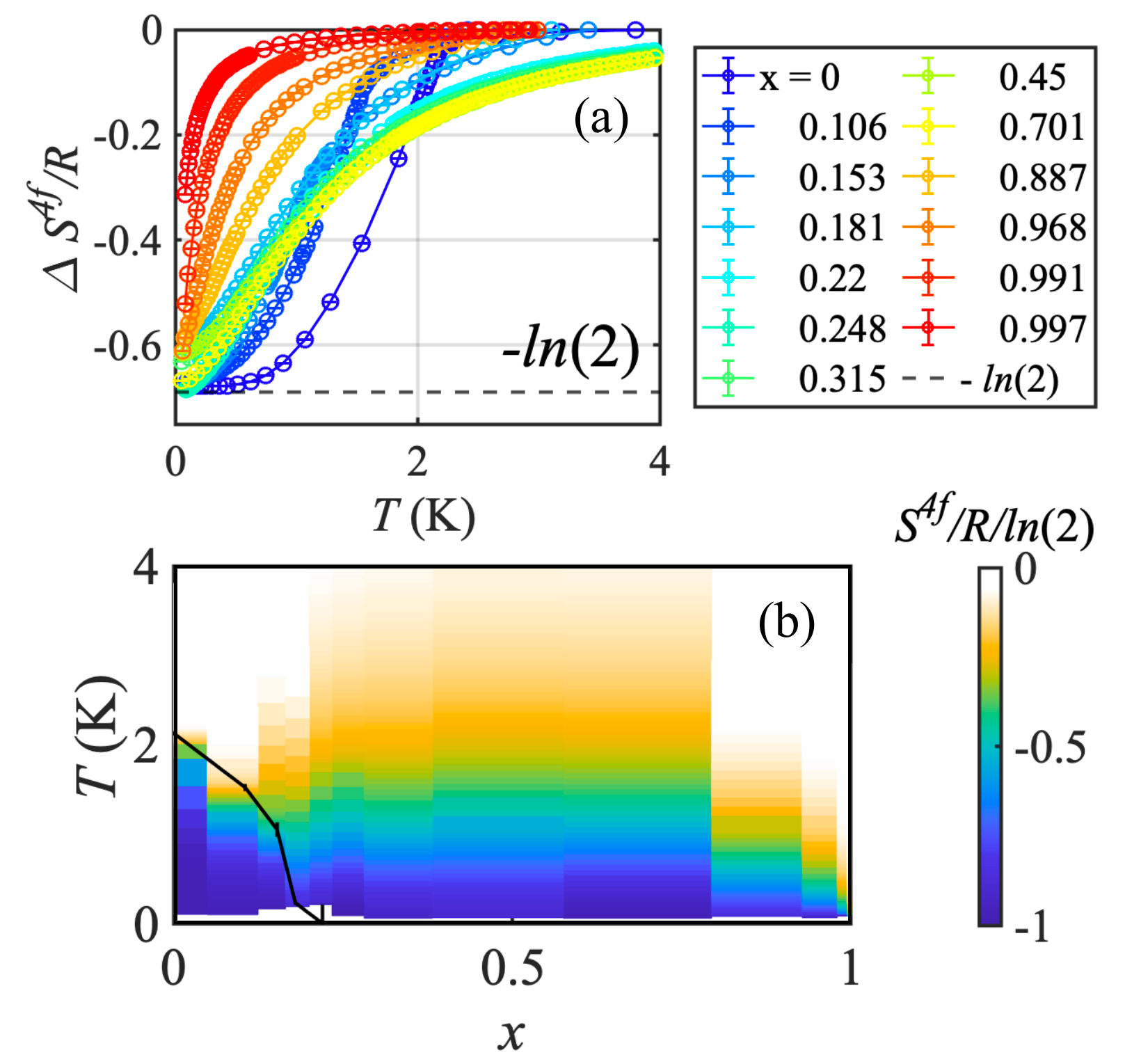}
	\caption{(a) Entropy change $\Delta S^{4f}$ normalized by $R$ as a function of $T$ at selected $x$ (see text). The dashed line shows the value of $-\ln(2)$. (b) Color map of $\Delta S^{4f}$ in the $T-x$ plane. The solid black line indicates the extracted ferroquadrupolar transition temperature $T_Q$.}\label{fig-4} 
\end{figure}

\section{\NoCaseChange{Strain distribution in \ce{Tm_{1-x}Y_xVO4}}}


In this section, we focus on a quantitative analysis of the evolution of the gap distribution $\rho(\Delta)$ with $x$. The distribution of gaps in a locally disordered two-level system was originally proposed in explaining the specific heat of spin glasses \cite{phillips1987two}. A continuous distribution of splittings of the ground-state doublet has also been reported in different pyrochlore compounds \cite{ Duijn2005, tachibana2007heat, martin2017disorder}. Our analysis builds on these and similar perspectives, and aims to uncover significant trends affecting the variation of $\rho(\Delta)$ with composition in \ce{Tm_{1-x}Y_{x}VO4}. 

Given a known PDF that represents the distributions of the gaps $\rho(\Delta)$, the total heat capacity can be obtained from the sum of individual Schottky anomalies associated with each $\Delta$:

\begin{equation} \label{eqn}
	\bar{C}_p^{4f}(T) = N \int_{0}^{\infty}{d\Delta \rho(\Delta)\frac{2\Delta^2}{T^2\cosh^2(\frac{2\Delta}{T})}}
\end{equation}


The inverse problem of determining $\rho(\Delta)$ from a measured heat capacity is much harder, as it is an inhomogeneous Fredholm integral equation of the first kind. The solution $\rho(\Delta)$  is then also subject to the constraints that it is nonnegative and normalized. Such a problem is ill-posed generally, with approximate numerical solutions that are sensitive to noise in the data \cite{markovskyIllposedProblems1988, yuanOverviewNumericalMethods2019}. In light of these difficulties, we take an alternative approach by circumventing the inverse problem altogether. Instead, we present a comparison of the measured heat capacity data to that which would be obtained from specific distributions, such as the 2D Gaussian distribution discussed below, which can provide a meaningful perspective on the actual PDF. Note that $\Delta$ here refers to half of the energy splitting between the two levels; we use this definition in the fitting and plotting procedure.

We consider the ground state doublet of $\ce{Tm^{3+}}$ in a \ce{Tm_{1-x}Y_xVO4} crystal environment. The disordered local strain environment results in a finite energy splitting via $B_{1g}$ and $B_{2g}$ symmetry distortions, where $\epsilon_{B_{1g}}=\epsilon_{x^2-y^2}=\epsilon_{xx}-\epsilon_{yy}$ and $\epsilon_{B_{2g}}=2\epsilon_{xy}$, with values that vary throughout the material. The magnitude of the total splitting $\Delta$ on any given site is proportional to the quadrature sum of the two strain components:  

\begin{equation} 
	    \Delta=\sqrt{(\eta_{1}\epsilon_{B_{1g}})^2+(\eta_{2}\epsilon_{B_{2g}})^2} \label{eq:Delta}
\end{equation}	
where $\eta_1$ and $\eta_2$ are the magneto-elastic coupling strength in the $B_{1g}$ and $B_{2g}$ channels. For simplicity of annotating the effect of the symmetry-breaking strain as two mutually transverse field components, we write:

\begin{equation} \label{eqn3}
    \begin{aligned}
    x&\equiv \eta_{1}\epsilon_{B_{1g}} \\
    y&\equiv \eta_{2}\epsilon_{B_{2g}} \\
    \Delta&\equiv\frac{1}{2}\Delta_{\mathrm{Schottky}}=\sqrt{x^2+y^2}
    \end{aligned}
	\end{equation}

We start with the empirical assumption that $x$ and $y$ are independent random variables that follow two Gaussian distributions $\rho_x(x)$ and $\rho_y(y)$, as one would expect from the central limit theorem. The width of the $\epsilon_{B_{1g}}$ and $\epsilon_{B_{2g}}$ distributions are $\sigma_a$ and $\sigma_b$. The expression for $\rho(\Delta)$ is given by the integral:
\begin{equation} \label{eqn4}
	    	 \rho(\Delta)=\int_{-\infty}^{\infty}{\int_{-\infty}^{\infty}{dx dy \rho_x(x)\rho_y(y) \delta(\Delta-\sqrt{x^2-y^2})} }
	\end{equation}
	
The solution for $\rho(\Delta)$ is then:
	
\begin{equation} \label{eq:2D_Gaussian_eqn}
    	     \begin{aligned}
    	 \rho(\Delta)&=\int_{-\Delta}^{\Delta}dx\frac{1}{\sigma_x\sqrt{2\pi}}e^{-\frac{1}{2}(\frac{x}{\sigma_x})^2}\frac{\Delta}{\sqrt{\Delta^2-x^2}}\\
      &\left[ \frac{1}{\sigma_y\sqrt{2\pi}}e^{-\frac{1}{2}(\frac{\sqrt{\Delta^2-x^2}}{\sigma_y})^2}+\frac{1}{\sigma_y\sqrt{2\pi}}e^{-\frac{1}{2}(\frac{-\sqrt{\Delta^2-x^2}}{\sigma_y})^2}\right] 
    \end{aligned}
\end{equation}

A detailed derivation of Eq. \ref{eq:2D_Gaussian_eqn} can be found in the Appendix \ref{app:derivation_of_2D_Gaussian}. We applied Eq. \ref{eq:2D_Gaussian_eqn} to Eq. \ref{eqn} to fit the experimental data. Since the total number of magnetic species $N$ is solely determined by the size and chemical composition of each sample, the two remaining free-fitting parameters are just $\sigma_a$ and $\sigma_b$.

\begin{figure}[t]
	\includegraphics[width = \columnwidth]{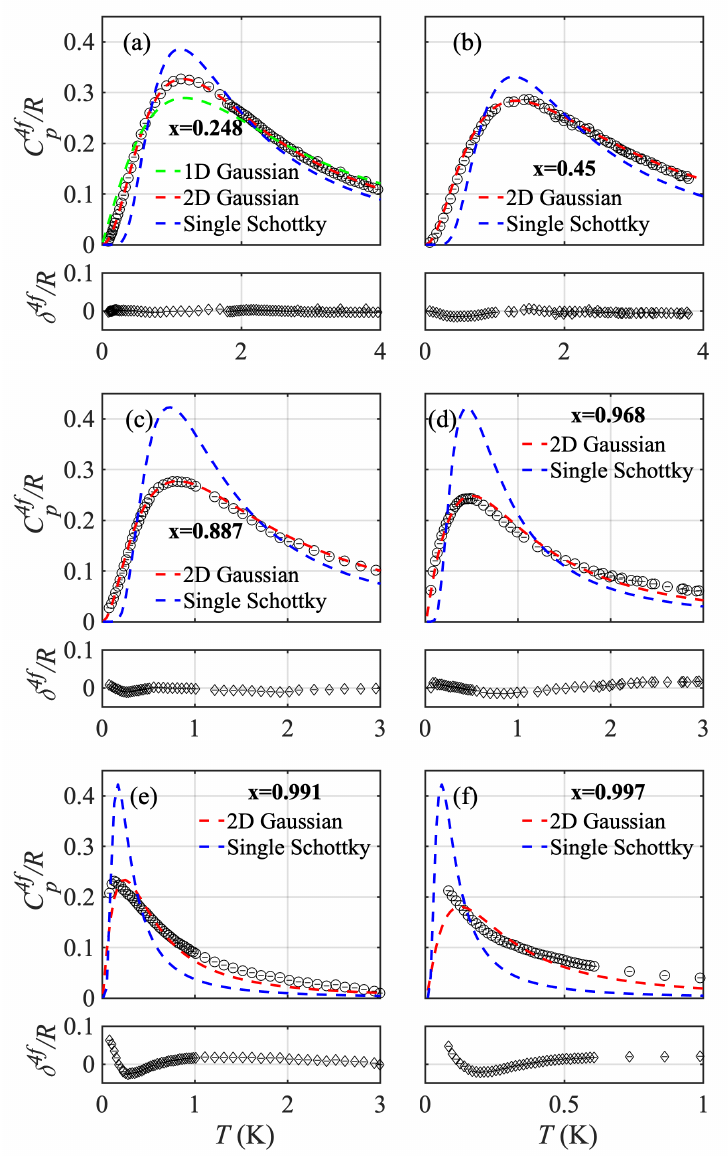}
	\caption{ (a)-(f) Upper panels plot the low-$T$ $4f$ heat capacity $C_p^{4f}$ of \ce{Tm_{1-x}Y_{x}VO4} with $x= 0.25, 0.45, 0.89, 0.97, 0.99$ and $0.997$. The blue dashed line shows a forced single-gap Schottky fitting with the data. The green dashed line in (a) represents the fitting result with Eq. \ref{eqn}, with $\rho(\Delta)$ given by a Gaussian PDF. The red dashed lines indicate fit to the 2D-Gaussian distribution in Eq. \ref{eq:2D_Gaussian_eqn}. Lower panels of (a-f) plot the residual $\delta^{4f}/R$ (fitting deviations) of each 2D Gaussian fit.  }\label{fig-5}
\end{figure}


Fig. \ref{fig-5} contrasts the fitting based on Eq. \ref{eq:2D_Gaussian_eqn} (red dashed lines) and forced fitting of a single Schottky anomaly (blue dashed lines) with $C^{4f}_p/R$ (black empty circles) for a few representative compositions $x$. The residual of the 2D Gaussian fits and the data $\delta^{4f}$ are also shown for each $x$ underneath each of the main panels. We note that the single Schottky anomaly scenario poorly fits the data for all $x$. In contrast, the 2D Gaussian distribution is found to describe the data very well for intermediate concentrations ($x=0.25 - 0.97$). However, with further increase in $x$, the distribution progressively deviates from the 2D Gaussian description, and becomes worst for the most dilute Tm concentrations, as can be seen in the large and systematic deviation from zero of $\delta^{4f}$ in the lower panels of Fig. \ref{fig-5}(e) and (f). 

\begin{figure}[t]
	\includegraphics[width =\columnwidth]{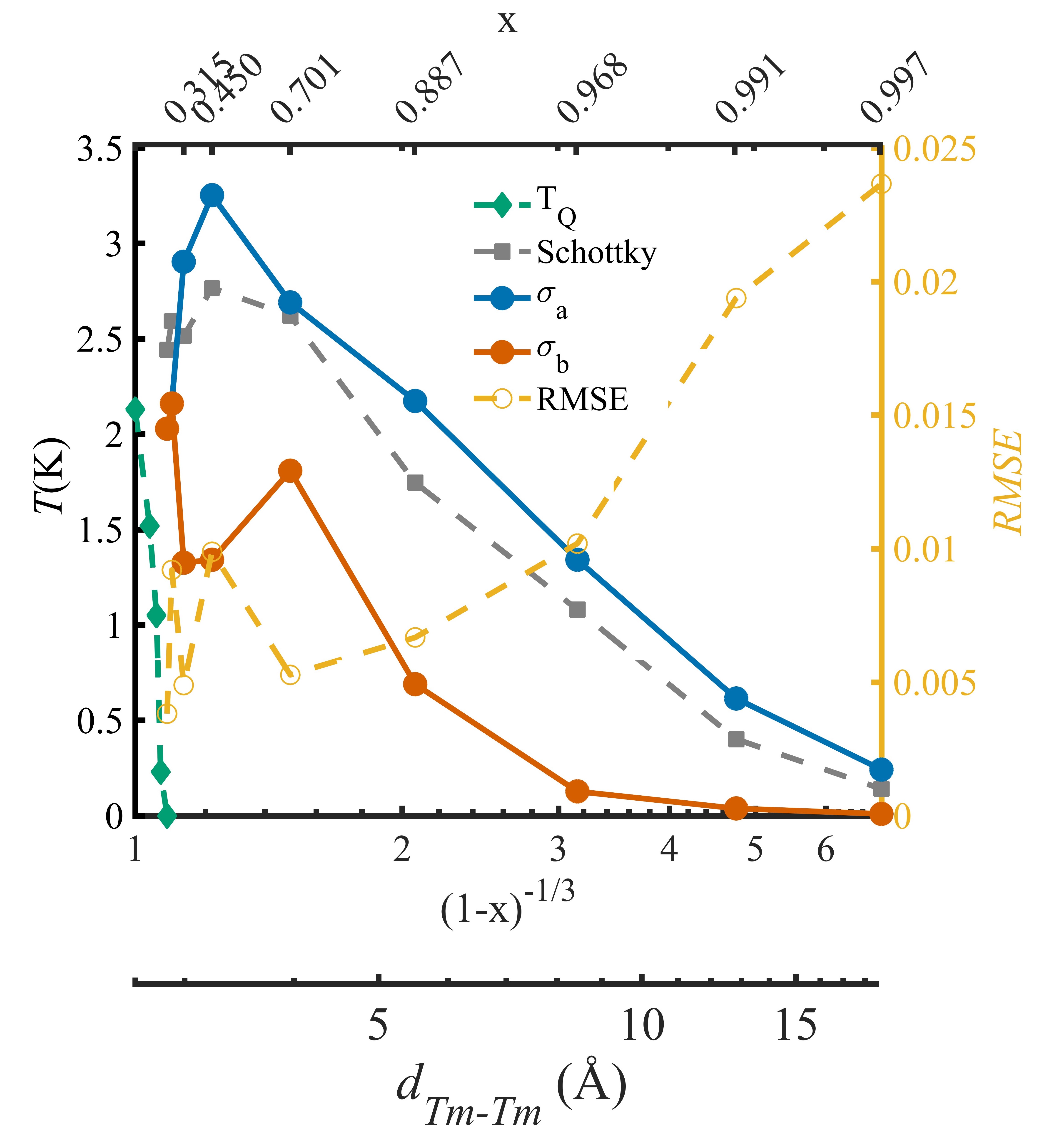}
	\caption{\label{fig-6} Evolution of the 2D-Gaussian distribution parameters plotted in the $T-x$ plane. The blue and red symbols represent $\sigma_a$ and $\sigma_b$ extracted from the fitting to Eq. \ref{eq:2D_Gaussian_eqn} (against the left axis). $T_Q$ is represented by the green diamond symbols. The gaps of a forced single Schottky anomaly fit are shown as gray symbols. The bottom axis is shown both as  $(1-x)^{-1/3}$ and  as $d_{Tm-Tm}$ (Wigner-Seitz radius of the \ce{Tm^{3+}}). The top axis indicates the $x$ value of each composition. The Root-Mean-Square-Error (RMSE) of each 2D Gaussian fitting is shown as yellow empty circles against the right axis. }
\end{figure}

Fig. \ref{fig-6} summarizes the evolution with composition of the fitting parameters, and the associated goodness of fitting (expressed as the root-mean-square error, denoted by RMSE), as a function of the average distance between Tm ions $d_{Tm-Tm}$. The abscissa is shown using a log scale, and is also expressed equivalently in terms of the composition as $(1-x)^{-1/3}$. The corresponding Y concentration $x$ is also labeled on the top axis. The two fitting parameters in the 2D Gaussian model, $\sigma_a$, and $\sigma_b$, which describe the width of the Gaussian distribution along the two-dimensional space, are shown as blue and red solid circles respectively. To provide a crude estimate of the characteristic energy scale associated with the strain-induced splitting, the result of the forced single-gap Schottky fit is also included (gray squares), though readers are reminded that this single-gap model provides a very poor fit, strongly suggesting the presence of a spatially inhomogeneous distribution of gaps. $T_Q$ is also shown on this plot as green diamonds. The yellow circles on the right axis indicate the RMSE of the respective 2D Gaussian fits. 

For compositions $x$ beyond approximately 50\%, the values of the gap obtained from the single Schottky fit and of the widths of the two independent Gaussian distributions, $\sigma_a$ and $\sigma_b$, rapidly decline in magnitude. This is in keeping with the general expectation that as the average distance between \ce{Tm^{3+}} ions increases (separated by increasingly undistorted YVO$_4$), the energy scale associated with lattice distortions gets progressively smaller, and the associated distribution gets narrower. However, $\sigma_b$ nearly vanishes for large enough $x$, resulting in physically unsound gap distributions $\rho(\Delta)$. Moreover, the systematic increase of RMSE with $x$ in the dilute limit reveals a significant deviation of the gap distribution from a 2D Gaussian description. This observation is robust against experimental uncertainties, including details of the subtraction of the phonon contribution to the heat capacity. This evidence for a progressive deviation of the local random strain distribution from a 2D Gaussian is our primary observation.

\section{\NoCaseChange{Discussion}}

According to the central limit theorem, for large enough sample sizes, the normalized sum of independent random variables tends towards a Gaussian PDF. In the present case, with potentially two independent variables (i.e. the two relevant components of the strain tensor, $\epsilon_{B_{1g}}$ and $\epsilon_{B_{2g}}$ experienced at each \ce{Tm^{3+}} site) a truly random distribution with no correlations would yield a $\rho(\Delta)$ following a 2D Gaussian distribution (see Fig. \ref{fig-1}). Our observations reveal that for a wide range of intermediate concentrations, the heat capacity in \ce{Tm_{1-x}Y_{x}VO4} cannot be distinguished from such a double-Gaussian distribution. Consequently, it is tempting to assume that a random field approach is justified for describing such a system. However, the observation of substantial deviations from the 2D Gaussian distribution for dilute concentrations indicates that a more subtle perspective is appropriate, and indeed that strains arising from substitution should not be treated as uncorrelated. In the following, we propose a scenario for the observed crossover of $\rho(\Delta)$ from a double Gaussian distribution in the concentrated Tm regime to a significant deviation from that in the dilute Tm regime, which emphasizes the presence of strain correlations in both limits.

\begin{figure}
	\includegraphics[width = \columnwidth, keepaspectratio]{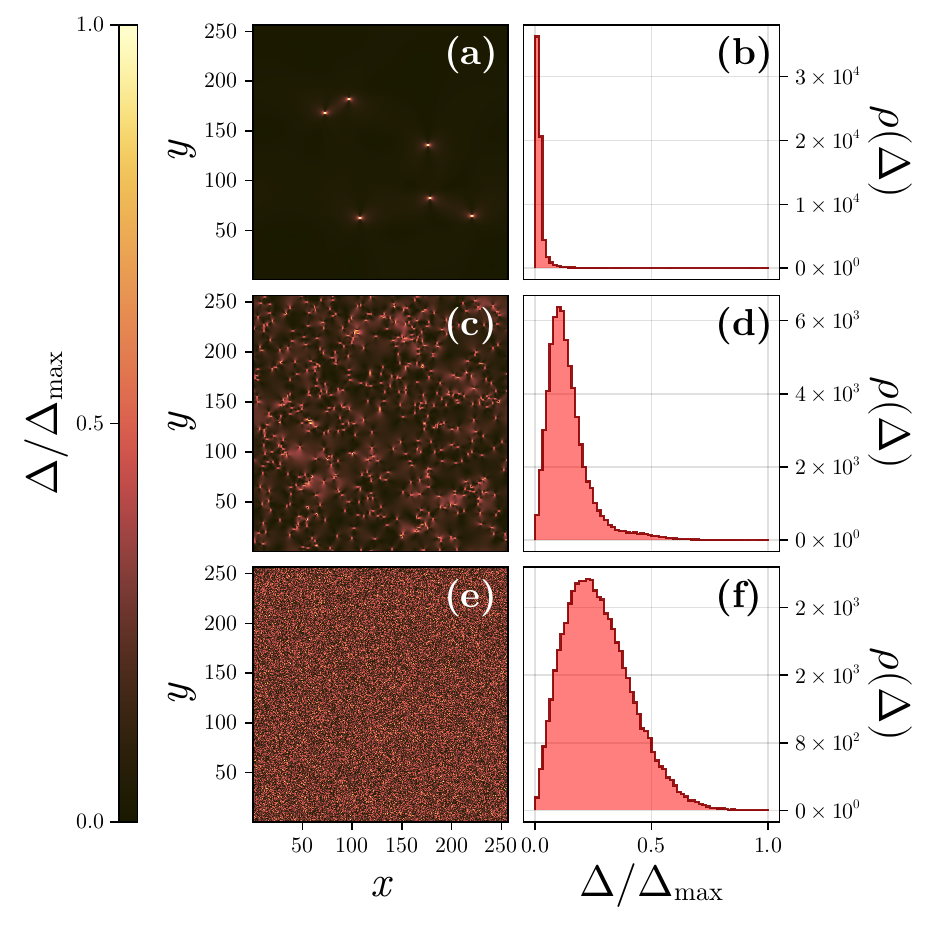} 
	\caption{Comparison of randomly scattered strain sources and their impact on the level-splitting distribution in \ce{Tm_{1-x}Y_{x}VO4}.  Each row shows a single realization of random level-splittings, $\Delta$,  in a two-dimensional periodic square lattice of size $256 \times 256$  on the left, and the corresponding measured distribution of $\Delta$, $\rho(\Delta)$, on the right. These splittings are normalized by each realization's maximum, $\Delta_{\max}$. In every case, the $B_{1g}$  and $B_{2g}$  strains are randomly drawn, and then Eq.  \ref{eq:Delta} is used to compute $\Delta$. The measured histograms are created with $256^2 = 65\,536$  values of $\Delta$. The first and second rows, plots \textbf{(a,b)}  and \textbf{(c,d)},  show the result of randomly placing defects throughout the simulation and computing the strains from elasticity theory, as explained in Appendix \ref{app:simulation_details}. In this case, the defects are edge dislocations.  \textbf{(a,b)}  represent a low concentration of defects (high $x$) whereas \textbf{(c,d)}  represent a high concentration of defects (low $x$). The bottom row, \textbf{(e,f)}, shows the case where both strains are Gaussian random and spatially uncorrelated, with a resulting histogram in the form in Eq.  \ref{eq:2D_Gaussian_eqn}.  As one introduces more crystalline defects into the system, one traverses the columns, and the distribution of level-splittings experienced by the $\ce{Tm^{3+}}$  ions approaches the 2D Gaussian.   \label{fig:dislocation_distribution_simulation}}
\end{figure}

Each Tm ion in \ce{Tm_{1-x}Y_{x}VO4} acts as a local reporter of the strain, with a splitting of the CEF ground state doublet that depends on the quadrature sum of the local values of $\epsilon_{B_{1g}}$ and $\epsilon_{B_{2g}}$. These strains originate from a variety of deviations from perfect crystallinity throughout the solid, including from extended defects such as dislocations, from residual strains that the crystal experiences from external stresses, and of course from the chemical substitution itself. To illustrate an important point, we briefly consider a thought experiment in which defects are introduced to TmVO$_4$ without diluting the Tm ions. We also imagine that the strains are small such that the response is strictly linear, and consequently that strains from various sources are simply superimposed. 

The outcome of this thought experiment is illustrated in Fig. \ref{fig:dislocation_distribution_simulation}. In the Figure, we modeled the effect of adding edge dislocations randomly throughout a two-dimensional periodic square lattice.  Using well-known expressions for the strain fields generated by a single disclocation, we compute $\epsilon_{B_{1g}}$  and $\epsilon_{B_{2g}}$ at each lattice site, from which we calculate the level-splitting experienced by each $\ce{Tm^{3+}}$ 
 ion. To avoid the formally diverging strain at the dislocation cores that emerge in linear elasticity, we place the dislocations at the centers of the square plaquettes whereas the two-level ions are positioned on each lattice site. For additional details of the model, please see Appendix \ref{app:simulation_details}. We choose edge dislocations as a tractable proof-of-concept to simulate the effects of long-range strain relaxation arising from a variety of actual defects. The real situation is of course more complex, with large local strains that presumably exceed the regime of linear response, but this simplification allows us to develop an important idea. 
 
 The strains associated with each defect relax as a power-law rather than exponential function of the distance from the defect, and hence there is no characteristic length scale for the relaxation. Nevertheless, there is an effective length scale beyond which the strains associated with any given defect become sufficiently small to be indistinguishable from any slowly varying background strain that arises not from intrinsic defects but, for instance, from small external stresses experienced by the material. This defines an effective radius over which the Tm ions are able to record the strains arising from defects. 

In the dilute limit, most $\ce{Tm^{3+}}$ ions report few strain sources, as there are only a few defects generating the internal strains[Figs. \ref{fig:dislocation_distribution_simulation}(a,b)]. Thus, the majority of the system is essentially undistorted, as indicated by the dark region in panel (a). It is clear that the aggregate strains on adjacent sites are highly correlated, as they are dominated by the strains generated by the defects. Although the locations of the defects are random, the number of defects within the ``sensing radius'' of any given ion is small enough that the central limit theorem does not apply.

We now imagine increasing the concentration of defects within this hypothetical system [Figs. \ref{fig:dislocation_distribution_simulation}(c,d)]. The defects are positioned randomly, so there is no correlation between their locations.  Since strains are correlated over long distances in the dilute limit, there must still be significant correlations between strains on adjacent sites even in the more concentrated regime. Nevertheless, as the concentration of defects is progressively increased, any given Tm site experiences strains arising from an increasingly large number of defects within its sensing volume, which, in the elastic limit, are simply added to give the total (local) strain that the site experiences. Since the dominant effect of the defect locations is random (uncorrelated), the distribution $\rho(\Delta)$ then starts to crossover into the 2D Gaussian regime as the central limit theorem locally begins to dominate the strain statistics in the $B_{1g}$  and $B_{2g}$ strain channels. It is important to emphasize, meanwhile, that one can obtain a $\rho(\Delta)$ distribution that approximates a 2D Gaussian even if the strains are correlated between sites. The extreme limit of short-ranged behavior would be realized only in samples with a high density of defects, as shown in Figs. \ref{fig:dislocation_distribution_simulation}(e,f).

Thus, within this thought experiment, a cross-over is anticipated from a regime in which clear signatures of strain correlations are apparent (in the dilute limit) to a regime in which the randomness of the defect locations dominates the strain distribution, and for which a Gaussian distribution of gaps is a reasonable approximation. The concentration of defects at which this crossover occurs is determined by the interplay of several key length scales: the effective length scale on which strains relax, the distance between defects, and the distance between Tm ``reporters". Signatures of ``randomness" only become apparent when the number of defects and the number of reporters within a typical relaxation volume become sufficiently large for the central limit theorem to apply.  

The situation in \ce{Tm_{1-x}Y_{x}VO4} is slightly more complex than the above thought experiment, but not in any qualitative way. The substitution of Y for Tm yields a progressive increase in the distance between Tm ``reporters", while also affecting the strain landscape. Similarly, the length scale over which strains relax might itself depend on the concentration of impurities. The interplay between length scales is still apparent, though, and dilute versus concentrated regimes with different qualitative behaviors can still be anticipated.   

 In addition to the above, we note that the constitutive elasticity relations (i.e. the relations between stress and strain, which at the level of linear response are governed by the elastic stiffness tensor) necessarily imply that some arbitrary stress state at a given site is a linear superposition of the various strain components at that site. Therefore, if the stress state is fixed by sources of structural disorder in the material, the six strain components cannot all be taken as independent of one another. Hence, $\epsilon_{B_{1g}}$ and $\epsilon_{B_{2g}}$ on any given site are correlated in some specific way determined by the elastic distortion of the crystal. We emphasize that this is in addition to the correlations that exist for the same strain component between adjacent sites.

\section{\NoCaseChange{Conclusions}}
Substitution of Y for Tm in \ce{Tm_{1-x}Y_{x}VO4} rapidly suppresses the long-range quadrupole order that is found for the pristine ``parent'' compound TmVO$_4$. Splitting of the CEF non-Kramers groundstate doublet of each Tm ion acts as a local reporter of the quadrature sum of $\epsilon_{B_{1g}}$ and $\epsilon_{B_{2g}}$ strains on each site. Our heat capacity data reveal that the splitting is not uniform, but is described by some probability distribution. While the heat capacity data are not amenable to an inverse analysis (in which the distribution is determined from the observed heat capacity), the data can nevertheless be compared to the simple expectations of a 2D Gaussian distribution that would otherwise be expected if the strains were truly random (uncorrelated). Significantly, for dilute concentrations the heat capacity cannot be accounted for by such a distribution, revealing for the first time clear evidence of strain correlations \textit{between} sites and between different strain components at the \textit{same} site. For larger concentrations, however, the heat capacity cannot be distinguished from that which would be obtained from a random distribution. The cross-over between these regimes can be understood in terms of the progressive change in the number of defects within the sensing volume of a given Tm `reporter' as the concentration of impurities $x$ is changed. In the concentrated regime, even though strains are correlated between sites, the defects that generate the strains themselves are randomly distributed. Hence, the total strain on each site is reasonably described by a Gaussian distribution when the number of defects within the sensing volume of a reporter ion is large enough for the central limit theorem to apply.  

Our results demonstrate that the strains, which arise from chemical substitution in \ce{Tm_{1-x}Y_{x}VO4}, cannot be regarded as uncorrelated random fields. Consequently, theoretical treatment of this problem (i.e. treatment of the problem of quadrupole moments that interact in the presence of non-uniform longitudinal and transverse effective fields that arise from local chemical substitution) cannot rely solely on the results of the random field Ising model. Instead, one must consider the subtle effects of strain correlations that can occur between different sites for the same symmetry channel, as well as between different symmetry channels, either between different sites or the same one. This problem is not restricted to just \ce{Tm_{1-x}Y_{x}VO4}. Any electronic nematic system for which chemical substitution is used to traverse the associated phase diagram falls into the same paradigm. Our results therefore motivate in-depth theoretical efforts to model the rich problem of correlated ``random'' fields on Ising electronic nematic order.

\section{\NoCaseChange{Acknowledgements}}
We thank Brad Ramshaw, Patrick Hollister, J\"{o}rg Schmalian, Mingde Jiang, and Evan Reed for fruitful discussions. Low-temperature heat-capacity measurements performed at Stanford University were supported by the Air Force Office of Scientific Research Award FA9550-20-1-0252, using a cryostat acquired with award FA9550-22-1-0084. Crystal-growth experiments were supported by the Gordon and Betty Moore Foundation Emergent Phenomena in Quantum Systems Initiative Grant GBMF9068. M.P.Z. was also partially supported by a National Science Foundation Graduate Research Fellowship under grant number DGE-1656518. L.Y. acknowledges support from Marvin Chodorow Postdoctoral Fellowship at the Department of Applied Physics, Stanford University. The theoretical work at the University of Minnesota (W.J.M. and R.M.F.)  was supported by the U.S. Department of Energy, Office of Science, Basic Energy Sciences, Materials Science and Engineering Division, under Award No. DE-SC0020045. DFT calculations by Y.Z. were supported by NSF DMR-1455050.  

\begin{appendices}
\section{\NoCaseChange{Determination of chemical composition via magnetic susceptibility} \label{app:chem_composition}}
The magnetic susceptibility ($\chi$) of \ce{TmVO4} along its $c$-axis exhibits a Curie-Weiss behavior down to $2K$ \cite{COOKE1972265}, which is not frequency-dependent in the absence of an external magnetic field \cite{Bleaney1980}. We observed the same $\chi$ behavior in \ce{(Y,Tm)VO4} above $T_Q$ (if it exists) owing to the identical crystal field environments throughout the doping series. Consequently, to accurately identify the chemical composition of samples that contain the lowest Tm percentages, we compare their magnetic susceptibility with the calculated specific magnetic susceptibility. In zero-field conditions, the calculated magnetic susceptibility can be expressed as:

\begin{equation} \label{eqn6}
	    	\chi =\frac{\frac{1}{4}\mu_0N g_C^2\mu_B^2}{k_B(T-T_{\theta})}
	\end{equation}

$\mu_0$ is the vacuum magnetic permeability. $N$ is number of \ce{Tm^{3+}} ions. $g_C=10.22$ is the g-factor of $\ce{Tm^{3+}}$ along the c-axis. $\mu_B$ is the Bohr magneton. $T_{\theta}$ is the Curie temperature. We fit the equation above with the measured ac magnetic susceptibility as:
\begin{equation} \label{eqn7}
	    	\chi =\frac{A}{k_B(T-T_{\theta})}+\chi_0
	\end{equation}

Three fitting parameters are $A=\frac{\mu_0 g_c^2 \mu_B^2 N}{4k_B}$ as the Curie constant, $T_{\theta}$ as the Curie temperature, and $\chi_0$ as the measured constant magnetic background from non-magnetic species (mostly from \ce{YVO4} lattices). 

The ac magnetic susceptibility measurement was performed with a Magnetic Properties Measurement System (MPMS), with a drive frequency of 75.7Hz. The value of $A$ takes the average of all the fitting results from 2K to different $T_{max}$ values. Finally, we obtain $x$ in the main text calculated from $N$, molar mass $M$, and the sample mass $m$.

\section{\NoCaseChange{{Specific Heat Model and Determination of $C_p^{4f}$}} \label{app:specific_heat_determination}}

In this section, we discuss the method of finding and subtracting the phonon background and obtaining the heat capacity per \ce{Tm} site ($C_p^{4f}$) from the heat capacity measurement. We assume the measured heat capacity is only the sum of the phonon contribution and the Schottky anomaly of the $4f$ degree of freedom:

  \begin{equation}\label{eqn8}
  \begin{aligned}
    C_p^m &= C_p^{Debye}+C_p^{Schottky}\\
    &=C_p^{Debye}+(1-x)C_p^{4f}
 \end{aligned}
\end{equation}

Where $x$ represents the content of $Y^{3+}$. The three-dimensional Debye model describes the phonon contribution of heat capacity. The fit equation is:

\begin{equation} \label{eqn9}
    C_p^{Debye}=9N k_B(\frac{T}{T_{eff}})^3\int_{0}^{\frac{T}{T_{eff}}}dx\frac{x^4e^x}{(e^x-1)^2}
	\end{equation}
 
The $C_p^{Debye}$ refers to the contribution of phonons to the total heat capacity. $T_{eff}$ is a fitting parameter that represents the effective Debye temperature. It is important to note that the fitting method used in the Debye Model with a limited temperature range is empirical and does not take into account the anisotropic phonon dispersion in the \ce{Y_{x}Tm_{1-x}VO4} series. Therefore, we use $T_{eff}$ to differentiate it from the experimentally determined $T_{Debye}$ that was reported elsewhere\cite{Denisova2015}.

We applied Eq.\ref{eqn3} to heat capacity data from $10$ to $25K$, where the Schottky contribution is negligible. 

To obtain $\Delta C_p^{Schottky}$, we calculated the phonon background at lower temperatures (shown as green and red dashed lines in Fig.\ref{fig-9}), using $T_{eff}$ obtained from Debye fitting. The phonon background below $T=0.65K$ was extrapolated using the low-temperature $T^{3}$ limit of the Debye model:

\begin{equation} \label{eqn10}
    C_p^{Debye}=36N k_B(\frac{T}{T_{eff}})^3
	\end{equation}
	
The fit result based on the specific heat model is plotted in Fig. \ref{fig-9} (j).

\begin{figure}
  \includegraphics[width=\linewidth]{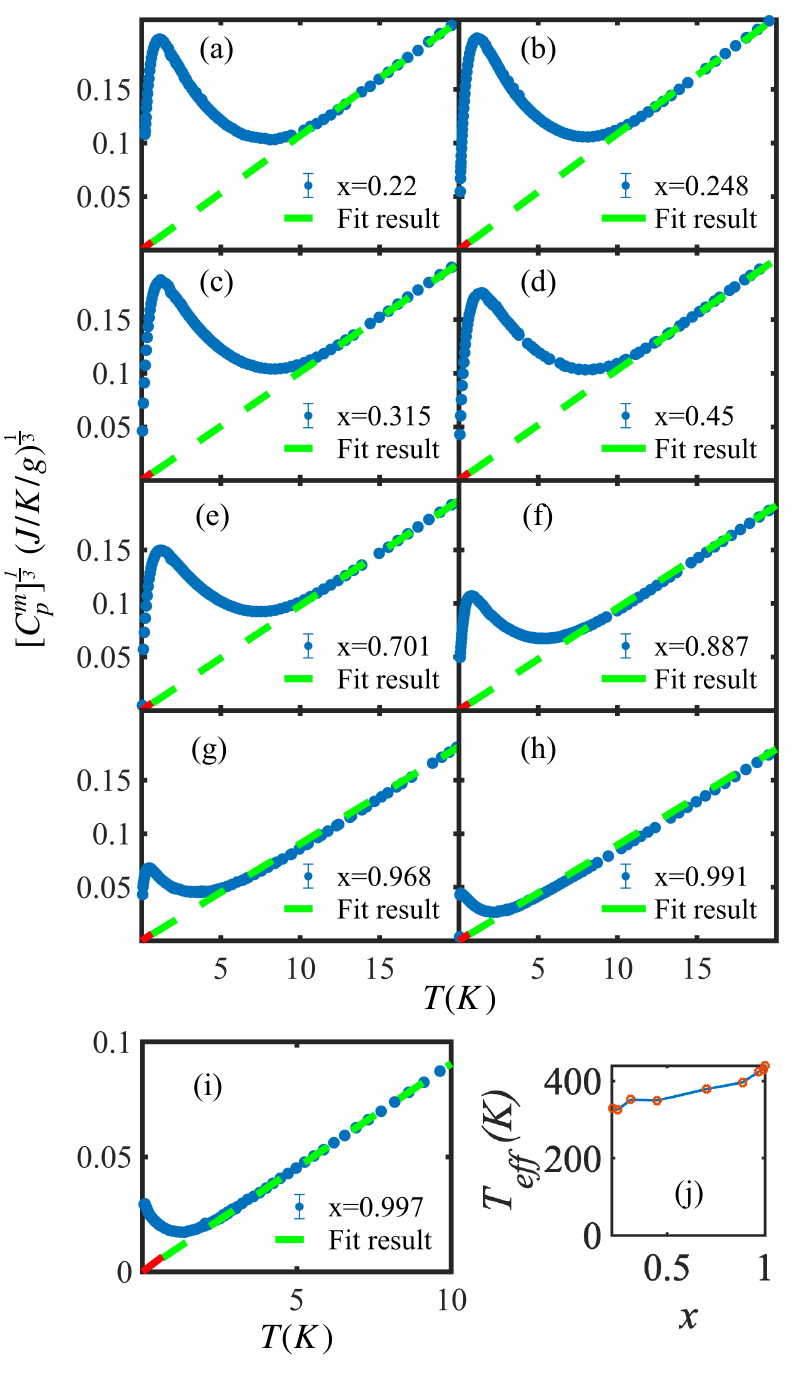}
  \caption{(a)-(i) Fitting result of the heat capacity data in the fitting range from 10K to 25K. The dashed curve plots the fitting equation based on the Debye Model. The green (red) dashed line region is based on the calculation of the Debye model (expansion of the low-temperature limit) extrapolated from the fit result. (j) The effective Debye Temperature $T_{eff}$ v.s. $x$ obtained from the fitting.}
  \label{fig-9}
\end{figure}

\section{\NoCaseChange{Derivation of the 2D-Gaussian distribution in \ce{(Y,Tm)VO4}} \label{app:derivation_of_2D_Gaussian}}

The ground state doublet of ${\rm Tm}^{3+}$ affected by local strain is described by Eq.\ref{eqn4} and Eq.\ref{eq:2D_Gaussian_eqn} in the main text. We assume that $\rho_x(x)$ and $\rho_y(y)$ are two probability distribution functions corresponding to the strength of the $\epsilon_{B1g}$ and $\epsilon_{B2g}$  respectively. Qualified by the uncorrelated random-field assumption, both of the functions are the Gaussian Distribution with a mean value centered at $0$. The probability distribution function of the quadrature sum $\Delta$ can be written as:

\begin{equation} \label{eqn11}
	    	 \rho(\Delta)=\int_{-\infty}^{\infty}{\int_{-\infty}^{\infty}{dx dy \rho_x(x)\rho_y(y) \delta(f(\Delta))} }
	\end{equation}
	
where $f(\Delta)$ is:

\begin{equation} \label{eqn12}
	    	 f(\Delta)=\Delta-\sqrt{x^2-y^2}
	\end{equation}	
	
Therefore, we can now solve for $\rho(\Delta)$ as: 
	
\begin{equation} \label{eqn13}
    \begin{aligned}
    	 \rho(\Delta)=&\int_{-\infty}^{\infty}{\int_{-\infty}^{\infty}{dx dy  \rho_x(x)\rho_y(y) \delta(\Delta-\sqrt{x^2-y^2}})} \\
       =&\int_{-\infty}^{\infty}dx{ \rho_x(x)[\int_{-\infty}^{\infty}{dy \rho_y(y) \delta(\Delta-\sqrt{x^2-y^2})}}]
    \end{aligned}
\end{equation}
The identity of a delta function implies:

\begin{equation} \label{eqn14}
	\delta[g(y)]=\sum_i \frac{\delta(y-y_i)}{|g'(y_i)|}
	\end{equation}
	
where $m_i$ are roots of the function $g(y)$. 
\begin{equation} \label{eqn15}
\begin{aligned}
	g(y)&=\Delta-\sqrt{x^2-y^2}\\
	y_i&=\pm \sqrt{\Delta_i^2-x^2}\\
	g'(y_i)&=\frac{\partial(\Delta-\sqrt{x^2-y^2})}{\partial{(\pm\sqrt{\Delta_i^2-x^2})}}=\pm\frac{\sqrt{\Delta_i^2-x^2}}{\Delta}
	\end{aligned}
	\end{equation}

Therefore, we now apply Eq.\ref{eqn14} to Eq.\ref{eqn13} and obtain:

\begin{equation} \label{eqn16}
    \begin{aligned}
    	 \rho(\Delta)&=\int_{-\Delta}^{\Delta}dx\rho_x(x)\frac{\Delta}{\sqrt{\Delta^2-x^2}}\\
      &[\rho_y(\sqrt{\Delta^2-x^2})+\rho_y(-\sqrt{\Delta^2-x^2})]
    \end{aligned}
	\end{equation}

Then, we apply two Gaussian distributions with widths $\sigma_x$ and $\sigma_y$ respectively, and calculate the analytical solution of the $\rho(\Delta)$:

\begin{equation} \label{eqn17}
    \begin{aligned}
    	 \rho_x(x)&=\frac{1}{\sigma_x\sqrt{2\pi}}e^{-\frac{1}{2}(\frac{x}{\sigma_x})^2}
    	 \\
    	 \rho_y(y)&=\frac{1}{\sigma_y\sqrt{2\pi}}e^{-\frac{1}{2}(\frac{y}{\sigma_y})^2}
    \end{aligned}
\end{equation}
	
The analytical solution for $\rho(\Delta)$ is the Eq.\ref{eq:2D_Gaussian_eqn} of the main text.

\section{\NoCaseChange{Modeling correlated strains from dislocations} \label{app:simulation_details}}

In this section, we describe the numerical process used to generate the level-splitting realizations shown in Fig. \ref{fig:dislocation_distribution_simulation}. We use the same procedure shown in \cite{Meese2024}. To best compare conventional short-ranged quenched disorder fields with long-ranged elastic strains generated by defects, we use the longest-ranged fundamental crystalline defect typically observed in equilibrium, the dislocation \cite{landauTheoryElasticity1970, muraMicromechanics}. Dislocations are topological defects with a charge known as the Burgers vector, which we denote as $\boldsymbol{b}$ \cite{chaikinLubensky}.  The strains are well known to decay as $1/r$ from the core of a straight dislocation, where $r$  is the radial distance to the dislocation line \cite{landauTheoryElasticity1970, muraMicromechanics}. In brief, we employ an ensemble of random, uncorrelated, edge dislocations to generate the $B_{1g}$ and $B_{2g}$  components of the strain tensor.  We use linear elasticity theory then to superimpose the strains from each dislocation within the ensemble at a given site. From there it follows that the level-splitting, $\Delta$,  is the quadrature sum of the net strains at each site. 

 In isotropic media, the displacement vector $\boldsymbol{u}$ induced by a single straight edge dislocation centered at the origin, oriented along the $z$-axis, and endowed  with Burgers vector $\boldsymbol{b}$  is known to be \cite{landauTheoryElasticity1970, muraMicromechanics}  
 \begin{align}
          u_x(\boldsymbol{r}; \boldsymbol{b}) &= \frac{b_x \phi}{2\pi} + \frac{1-2\nu}{4\pi(1-\nu)}b_y\ln r \nonumber
     \\
     &\quad + \frac{1}{8\pi (1 - \nu)}\left( b_x\sin 2\phi -b_y\cos 2\phi \right),
 \end{align}
 \begin{align}
     u_y(\boldsymbol{r};\boldsymbol{b}) &= \frac{b_y \phi}{2\pi} - \frac{1-2\nu}{4\pi(1-\nu)}b_x\ln r \nonumber
     \\
     &\quad - \frac{1}{8\pi (1 - \nu)}\left( b_x\cos 2\phi + b_y\sin 2\phi \right),
 \end{align}
 where $\boldsymbol{r}=(x,y)=r(\cos\phi,\sin\phi)$  and $\nu$  is the Poisson ratio. From these expressions,  $\epsilon_{B_{1g}}$  and $\epsilon_{B_{2g}}$ are obtained by differentiating the displacement vector as 
 \begin{align}
     \epsilon_{B_{1g}}(\boldsymbol{r};\boldsymbol{b}) &= \partial_x u_x - \partial_y u_y = -\frac{\left( \boldsymbol{b} \cdot \boldsymbol{\hat{r}} \right) \sin 2\phi }{2\pi(1-\nu)r},
     \\
     \epsilon_{B_{2g}}(\boldsymbol{r};\boldsymbol{b}) &= \partial_x u_y + \partial_y u_x = \frac{\left( \boldsymbol{b} \cdot \boldsymbol{\hat{r}} \right) \cos 2\phi }{2\pi(1-\nu)r}.
 \end{align}
 These functions are used to compute the strain generated by a dislocation at every lattice site in a two-dimensional periodic square grid with properly enforced periodic boundary conditions \cite{Meese2024}. To avoid the singular behavior of the strain at the dislocation cores, we place them at the centers of the square plaquettes.
 
 The total strain in the $a\in\{B_{1g},B_{2g}\}$ irreducible representation is then computed from a set of  $N_d$ dislocations as 
 \begin{align}
     \epsilon_{a}(\boldsymbol{r}) = \sum_{\gamma = 1}^{N_d} \epsilon_{a}\left( \boldsymbol{r} - \boldsymbol{r}_\gamma; \boldsymbol{b}_\gamma \right),
 \end{align}
 from which it follows that the level-splittings are 
 \begin{align}
     \Delta(\boldsymbol{r}) = \sqrt{\eta_{1}^2\epsilon_{B_{1g}}^2(\boldsymbol{r}) + \eta_{2}^2\epsilon_{B_{2g}}^2(\boldsymbol{r})}.
 \end{align}
 For simplicity, we have maintained the isotropic approximation, and have set the magneto-elastic couplings to be equal: $\eta_{1} = \eta_{2}=\lambda $.  In Figure \ref{fig:dislocation_distribution_simulation}, we only consider ratios of $\Delta(\boldsymbol{r})$   to its maximum value in the lattice, $\Delta_{\max}$,  which removes the numerical dependence on $\lambda$ and $\nu$ without losing the qualitative features of a long-ranged correlated random strain field.
 
Each realization of random strain is created via an ensemble of $N_d$ random dislocations. Equilibrium distributions of dislocations must be  ``dislocation neutral"  \cite{kleinertGaugeFieldsSolids1989}.  Thus, we impose that the dislocations within each realization appear in pairs of equal and opposite topological charges.  This is the only nonlinear effect we model, however. Thus, dislocations were assumed to be nearly uncorrelated, up to the condition that the total Burgers vector vanishes and no two dislocations occupy the same plaquette. Up to these restrictions, the edge dislocations were created by sampling the Burgers vectors uniformly from the lattice vectors $\left\{ \pm \boldsymbol{\hat{x}}, \pm\boldsymbol{\hat{y}} \right\}$ and uniformly sampling the core locations to be at the center of one of the $L^2$  square plaquettes.

The calculation was performed with the \textit{Julia} programming language \cite{bezansonJuliaFreshApproach2017}
 and Fig. \ref{fig:dislocation_distribution_simulation} was created using the \textit{Makie} plotting package \cite{danischMakieJlFlexible2021}.
\end{appendices}

\end{document}